\documentclass{PoS}
\usepackage{amsmath}

\newcommand{\de}{\ensuremath{\delta} }
\newcommand{\eps}{\ensuremath{\epsilon} }
\newcommand{\ka}{\ensuremath{\kappa} }
\newcommand{\la}{\ensuremath{\lambda} }
\newcommand{\cD}{\ensuremath{\mathcal D} }
\newcommand{\cDbar}{\ensuremath{\overline{\mathcal D}} }
\newcommand{\Ibb}{\ensuremath{\mathbb I} }
\newcommand{\cN}{\ensuremath{\mathcal N} }
\newcommand{\cO}{\ensuremath{\mathcal O} }
\newcommand{\cP}{\ensuremath{\mathcal P} }
\newcommand{\cQ}{\ensuremath{\mathcal Q} }
\newcommand{\cU}{\ensuremath{\mathcal U} }
\newcommand{\cUbar}{\ensuremath{\overline{\mathcal U}} }
\newcommand{\lalat}{\ensuremath{\lambda_{\rm lat}} }
\newcommand{\muhat}{\ensuremath{\widehat \mu} }
\newcommand{\UN}{\ensuremath{\mbox{U(}N\mbox{)}} }
\newcommand{\SUN}{\ensuremath{\mbox{SU(}N\mbox{)}} }
\newcommand{\Uone}{\ensuremath{\mbox{U(1)}} }
\renewcommand{\Re}{\ensuremath{\mbox{Re}} }
\renewcommand{\Im}{\ensuremath{\mbox{Im}} }
\newcommand{\nn}{\nonumber}
\newcommand{\X}{\ensuremath{\!\times\!} }
\newcommand{\Tr}[1]{\ensuremath{\mbox{Tr}\left[ #1 \right]} }
\newcommand{\vev}[1]{\ensuremath{\left\langle #1 \right\rangle} }
\newcommand{\eq}[1]{Eq.~\ref{#1}}
\newcommand{\fig}[1]{Fig.~\ref{#1}}
\newcommand{\refcite}[1]{Ref.~\cite{#1}}

\title{Aspects of lattice $\cN = 4$ supersymmetric Yang--Mills}
\ShortTitle{Aspects of lattice $\cN = 4$ supersymmetric Yang--Mills}

\author{\speaker{David Schaich} \\
  Department of Physics, Syracuse University, Syracuse, NY 13244, United States\thanks{Permanent address} \\
  \mbox{Kavli Institute for Theoretical Physics, University of California, Santa Barbara, CA 93106, United States} \\
  Institut f\"ur Physik, Humboldt-Universit\"at zu Berlin, 12489 Berlin, Germany \\
  E-mail: \email{dschaich@syr.edu}
}

\abstract{ 
  Non-perturbative investigations of $\cN = 4$ supersymmetric Yang--Mills theory formulated on a space-time lattice have advanced rapidly in recent years.
  Large-scale numerical calculations are currently being carried out based on a construction that exactly preserves a single supersymmetry at non-zero lattice spacing.
  A recent development is the creation of an improved lattice action through a new procedure to regulate flat directions in a manner compatible with this supersymmetry, by modifying the moduli equations.
  In this proceedings I briefly summarize this new procedure and discuss the parameter space of the resulting improved action that is now being employed in numerical calculations.
}

\FullConference{The 33rd International Symposium on Lattice Field Theory \\ 14--18 July 2015 \\ Kobe, Japan}

\begin{document}
\setlength{\abovedisplayskip}{6 pt}
\setlength{\belowdisplayskip}{6 pt}
$\cN = 4$ supersymmetric Yang--Mills (SYM) is a particularly interesting gauge theory that plays important roles in holographic approaches to quantum gravity, investigations of the structure of scattering amplitudes, and the conformal bootstrap program.
It is also the only known four-dimensional theory admitting a lattice regularization that exactly preserves a subset of the supersymmetry algebra at non-zero lattice spacing~\cite{Kaplan:2005ta, Unsal:2006qp, Catterall:2007kn, Damgaard:2008pa, Catterall:2009it}.
This lattice construction provides a promising foundation for large-scale numerical investigations of $\cN = 4$ SYM that can in principle access non-perturbative couplings for arbitrary numbers of colors $N$.
(Other approaches to studying $\cN = 4$ SYM numerically include Refs.~\cite{Ishii:2008ib, Ishiki:2008te, Ishiki:2009sg, Hanada:2010kt, Honda:2011qk, Honda:2013nfa, Hanada:2013rga}.)

Even though the field of lattice $\cN = 4$ SYM is in its infancy, recent computations have provided the first ab~initio numerical results for quantities such as the static potential, to be confronted with perturbative and holographic predictions~\cite{Catterall:2014vka, Catterall:2014vga}. 
An exciting new development is the introduction of a procedure to regulate flat directions by modifying the moduli equations in a way that preserves the single exact supersymmetry at non-zero lattice spacing~\cite{Catterall:2015ira, Schaich:2015daa}.
This procedure results in an improved lattice action that exhibits dramatically reduced violations of supersymmetric Ward identities and much more rapid approach to the continuum limit.
This improved action has been implemented in our parallel software for lattice $\cN = 4$ SYM~\cite{Schaich:2014pda}, which we make publicly available to encourage independent investigations and the development of a lattice $\cN = 4$ SYM community.\footnote{{\tt\href{http://github.com/daschaich/susy}{http://github.com/daschaich/susy}}}

In this proceedings I briefly summarize the new procedure to regulate flat directions without breaking the exact supersymmetry, and discuss the parameter space of the resulting improved lattice action.
Preliminary results for the static potential and pfaffian phase from ongoing numerical calculations using the improved action were presented in \refcite{Schaich:2015daa}.
These investigations, as well as finite-size scaling and Monte Carlo renormalization group~\cite{Catterall:2014mha} analyses of the scaling dimensions of simple conformal operators, will soon be reported in future work.

\section*{Improved lattice action for $\cN = 4$ SYM} 
The underlying lattice action for $\cN = 4$ SYM is the direct discretization of the continuum action produced by topological twisting~\cite{Marcus:1995mq, Kapustin:2006pk},
\begin{align}
  S_{\rm formal} = \frac{N}{2\lalat} \sum_n \Tr{\cQ \left(\chi_{ab}(n)\cD_a^{(+)}\cU_b(n) + \eta(n) \cDbar_a^{(-)}\cU_a(n) - \frac{1}{2}\eta(n) d(n) \right)} \label{eq:S0} \\
                 -\frac{N}{8\lalat} \sum_n \Tr{\eps_{abcde}\ \chi_{de}(n + \muhat_a + \muhat_b + \muhat_c) \cDbar_c^{(-)} \chi_{ab}(n)},                                   \nn
\end{align}
with repeated indices summed and continuum gauge-covariant derivatives replaced by lattice finite-difference operators $\cD_a$~\cite{Catterall:2007kn, Damgaard:2008pa}.
All indices run from 1 through 5, corresponding to a discretization on the $A_4^*$ lattice of five linearly dependent basis vectors symmetrically spanning four space-time dimensions~\cite{Unsal:2006qp, Catterall:2014vga}.
At each lattice site $n$ the five complexified gauge links $\cU_a(n)$ contain both the gauge and scalar fields.
This results in an enlarged $\UN = \SUN \otimes \Uone$ gauge invariance, where $N$ is the number of colors.
In the continuum the U(1) sector decouples from observables in the SU($N$) sector, but this is not true at non-zero lattice spacing $a$.
\cQ is the twisted-scalar supersymmetry, whose closed subalgebra $\cQ^2 = 0$ is exactly preserved even for $a > 0$.

The equations of motion for the bosonic auxiliary field $d(n) = \cDbar_a^{(-)}\cU_a(n)$ determine the moduli space of the system.
The moduli space of the lattice theory matches that of the continuum theory~\cite{Catterall:2011pd}, and in particular possesses flat directions and exact zero modes that destabilize numerical computations.
These flat directions must be regulated in both the SU($N$) and U(1) sectors, generically requiring two deformations of the lattice action.
In Refs.~\cite{Catterall:2014vka, Schaich:2014pda, Catterall:2014vga}, these were added directly to \eq{eq:S0} to produce the unimproved action $S_{\rm unimp} = S_{\rm formal} + S_{\rm soft}$, where
\begin{equation}
  \label{eq:Ssoft}
  S_{\rm soft} = \frac{N}{2\lalat} \mu^2 \sum_n \sum_a \left(\frac{1}{N} \Tr{\cU_a(n) \cUbar_a(n)} - 1\right)^2 + \ka \sum_n \sum_{a < b} \left|\det \cP_{ab}(n) - 1\right|^2
\end{equation}
with two tunable auxiliary parameters $\mu$ and $\ka$.
The first ($\mu$) term is a scalar potential that regulates the SU($N$) flat directions and constrains $\vev{\Im\det\cP_{ab}}$, where $\cP_{ab} = \cP_{ba}^*$ is the oriented plaquette in the $a$--$b$ plane.
The U(1) phase of the links cancels out of $\Tr{\cU_a(n) \cUbar_a(n)}$ in this scalar potential, requiring the addition of the second ($\ka$) term to further constrain the plaquette determinant.

Both terms in \eq{eq:Ssoft} softly break the \cQ supersymmetry preserved by the underlying \eq{eq:S0}.
In the relevant region of parameter space the soft-breaking effects of the \ka term are much larger than those of the $\mu$ term~\cite{Catterall:2014vka}.
This motivated our recent development of a general method that can be applied to regulate flat directions in a manner compatible with the \cQ supersymmetry~\cite{Catterall:2015ira}.
The procedure produces modified auxiliary field equations of motion
\begin{equation}
  \label{eq:EoM}
  d(n) = \cDbar_a^{(-)}\cU_a(n) + G \cO(n)\Ibb_N,
\end{equation}
where $\cO(n)$ can be any gauge-invariant local bosonic operator and $G$ is a new auxiliary parameter.
The modified equations of motion are obtained by replacing
\begin{equation}
  \cQ\; \Tr{\eta(n) \left(\cDbar_a^{(-)}\cU_a(n)\right)} \longrightarrow \cQ\; \Tr{\eta(n) \left(\cDbar_a^{(-)}\cU_a(n) + G\cO(n)\Ibb_N\right)},
\end{equation}
in \eq{eq:S0}, introducing a manifestly $\cQ$-exact deformation.
The operator $\cO(n)$ is now constrained by a \cQ Ward identity,
\begin{equation}
  \label{eq:Ward}
  \vev{\sum_n \Tr{\cQ\; \eta(n)}} = \vev{\sum_n \Tr{d(n)}} = NG \vev{\sum_n \cO(n)} = 0,
\end{equation}
since the $\Tr{\cDbar_a^{(-)} \cU_a(n)}$ term in \eq{eq:EoM} vanishes upon summing over all lattice sites $n$.

We take $\cO(n) = \sum_{a \neq b} \left(\det\cP_{ab}(n) - 1\right) = 2\Re\sum_{a < b}\left(\det\cP_{ab}(n) - 1\right)$ in order to replace the \ka term in \eq{eq:Ssoft} that is responsible for the bulk of the soft \cQ supersymmetry breaking.
The corresponding Ward identity in \eq{eq:Ward} becomes $\sum_n \sum_{a \neq b} \vev{\det\cP_{ab}(n) - 1} = 0$, implying $\vev{\Re\det\cP_{ab}} = 1$. 
We still need to retain the $\mu$ term in \eq{eq:Ssoft}, to regulate flat directions in the SU($N$) sector and constrain $\vev{\Im\det\cP_{ab}}$.
This combination of a supersymmetric plaquette determinant deformation and a soft $\cQ$-breaking scalar potential defines the improved lattice action $S_{\rm imp}$ that we are now using in large-scale numerical calculations.

It may be possible to implement both of the necessary deformations in the $\cQ$-exact manner enabled by the new procedure summarized above.
In order to do so, however, we would have to make all the scalar potential and plaquette determinant terms non-negative at each lattice site so that they cannot cancel each other out and hence fail to regulate the flat directions.
We explored one way of accomplishing this in \refcite{Catterall:2015ira}, implementing the moduli equations
\begin{equation}
  \label{eq:overC}
  d(n) = \cDbar_a^{(-)}\cU_a(n) + B^2 \sum_a \left(\frac{1}{N} \Tr{\cU_a(n) \cUbar_a(n)} - 1\right)^2 \Ibb_N + G\sum_{a \neq b} \left|\det\cP_{ab}(n) - 1\right|^2 \Ibb_N,
\end{equation}
with auxiliary parameters $B$ and $G$. 
This choice turned out not to work, most likely because it imposes $15V$ non-trivial constraints for lattice volume $V$, rather than the single constraint of \eq{eq:Ward}.
Thus, although this over-constrained action $S_{\rm over}$ is exactly supersymmetric, lattice calculations employing it are generally unable to reach a supersymmetric vacuum.
Even so, it may be worthwhile to continue searching for more clever ways to further reduce soft supersymmetry breaking in numerical calculations.

\section*{Ward identities and the lattice parameter space} 
Our publicly available parallel software has been updated so that the user may employ any of the lattice actions discussed above ($S_{\rm formal}$, $S_{\rm unimp}$, $S_{\rm imp}$ or $S_{\rm over}$) by using appropriate values for the auxiliary parameters $\mu$, $\ka$, $G$ and $B$ in Eqs.~\ref{eq:Ssoft}, \ref{eq:EoM} and \ref{eq:overC}.\footnote{A compilation flag switches between the two different $G$ terms in \eq{eq:EoM} vs.\ \eq{eq:overC}.}
All of our ongoing calculations use the improved action by fixing $\ka = 0$ and $B = 0$ with non-zero values for $\mu$ and $G$.
These two couplings, in combination with the number of colors $N$, the lattice volume $L^4$ or $L^3\X N_t$ and the lattice 't~Hooft coupling $\lalat$, produce a somewhat complicated parameter space that takes some time to explore.
Here I summarize some results of that exploration and our resulting choices for the auxiliary parameters we are currently using.

To begin, recall that in the end we want to remove the two deformations in the improved action, in order to recover the full $\cN = 4$ SYM theory in the continuum limit $(a / L) \to 0$.
This motivates working with the smallest acceptable values of $\mu$ and $G$, extrapolating at least ${\mu \to 0}$.
If these parameters are made too small, however, the lattice calculations will exhibit instabilities, either confinement via U(1) monopole condensation or an excursion along a nearly flat direction (cf.~Figs.~1 and 5 in \refcite{Catterall:2014vka}).
The onset of these instabilities is quite sensitive to both the lattice volume and the 't~Hooft coupling.\footnote{We also observe some preliminary signs of sensitivity to $N$, but this appears less significant.}
As \lalat increases, either or both of $\mu$ and $G$ must be increased.
Conversely, these parameters can be decreased as $L$ increases towards the continuum limit.

At present we proceed by keeping $G$ fixed and scaling $\mu$ to produce constant $(\mu L)^2 / \lalat$.
The $\mu \propto 1 / L$ dependence can be motivated by thinking of $\mu$ as providing an effective mass for the scalar U(1) modes.
Scaling $\mu^2 \propto \lalat$ follows from the form of the scalar potential in \eq{eq:Ssoft}.
Finally, we fix $G$ both for simplicity and because the U(1) sector which it affects decouples from the SU($N$) target theory in the $(a / L) \to 0$ continuum limit, as we will see below~\cite{Catterall:2015ira}.

We monitor \cQ Ward identities to assess the amount of soft supersymmetry breaking and U(1) lattice artifacts.
For the improved action three such Ward identities are available: \eq{eq:Ward} fixes $\Re\det\cP = 1$; the structure of \eq{eq:S0} predicts the exact value of the bosonic action per lattice site $s_B = 9N^2 / 2$; and \cQ acting on $\cO \equiv \Tr{\eta \sum_a \cU_a \cUbar_a}$ produces the relative quantity
\begin{equation}
  \label{eq:bilin}
  \de \cQ \cO = \frac{\Tr{d \sum_a \cU_a \cUbar_a} - \Tr{\eta \sum_a \psi_a \cUbar_a}}{\sqrt{\Tr{d \sum_a \cU_a \cUbar_a}^2 + \Tr{\eta \sum_a \psi_a \cUbar_a}^2}},
\end{equation}
where the normalization factor in the denominator is slightly different than that used in \refcite{Catterall:2014vka}.
In addition we monitor the plaquette and the Polyakov loop $|L|$ constructed from the complexified gauge links (which are not unitarized).
$L$ is sometimes called the Maldacena loop, and since $\cQ |L| = 0$ it should be roughly $|L| \simeq 1$ for all $\lalat$.
The onset of instabilities generally leads to either $|L| \ll 1$ (for U(1) confinement) or $|L| \gg 1$ (for an excursion along a nearly flat direction).

\begin{table}[htp]
  \caption{\label{tab:continuum} Some observables from U(2) $L^4$ ensembles with $\lalat = 1$, $G = 0.05$ and $(\mu L)^2 / \lalat \approx 2.5$.}
  \centering
  \renewcommand\arraystretch{1.2}  
  \begin{tabular}{|ccc|ccccc|}
    \hline
    $L$ & $\mu$ & $G$   & Plaq.       & $|L|$       & $s_B / 18$    & $\Re\det\cP$  & $1 - \de\cQ\cO$ \\
    \hline
    16  & 0.10  & 0.05  & 1.0363(9)   & 0.918(113)  & 0.99986(4)    & 0.99864(19)   & 0.9962(5)       \\
    12  & 0.13  & 0.05  & 1.0285(12)  & 0.869(54)   & 0.99956(5)    & 0.99803(30)   & 0.9939(9)       \\
    8   & 0.20  & 0.05  & 1.0373(21)  & 1.147(71)   & 0.99877(13)   & 0.99507(52)   & 0.9837(19)      \\
    6   & 0.25  & 0.05  & 1.0565(56)  & 1.015(37)   & 0.99868(31)   & 0.98999(158)  & 0.9654(52)      \\
    4   & 0.40  & 0.05  & 1.0780(91)  & 1.118(33)   & 0.99381(107)  & 0.97875(494)  & 0.8848(237)     \\
    \hline
  \end{tabular}
\end{table}

\begin{figure}[bp]
  \centering
  \includegraphics[width=0.425\linewidth]{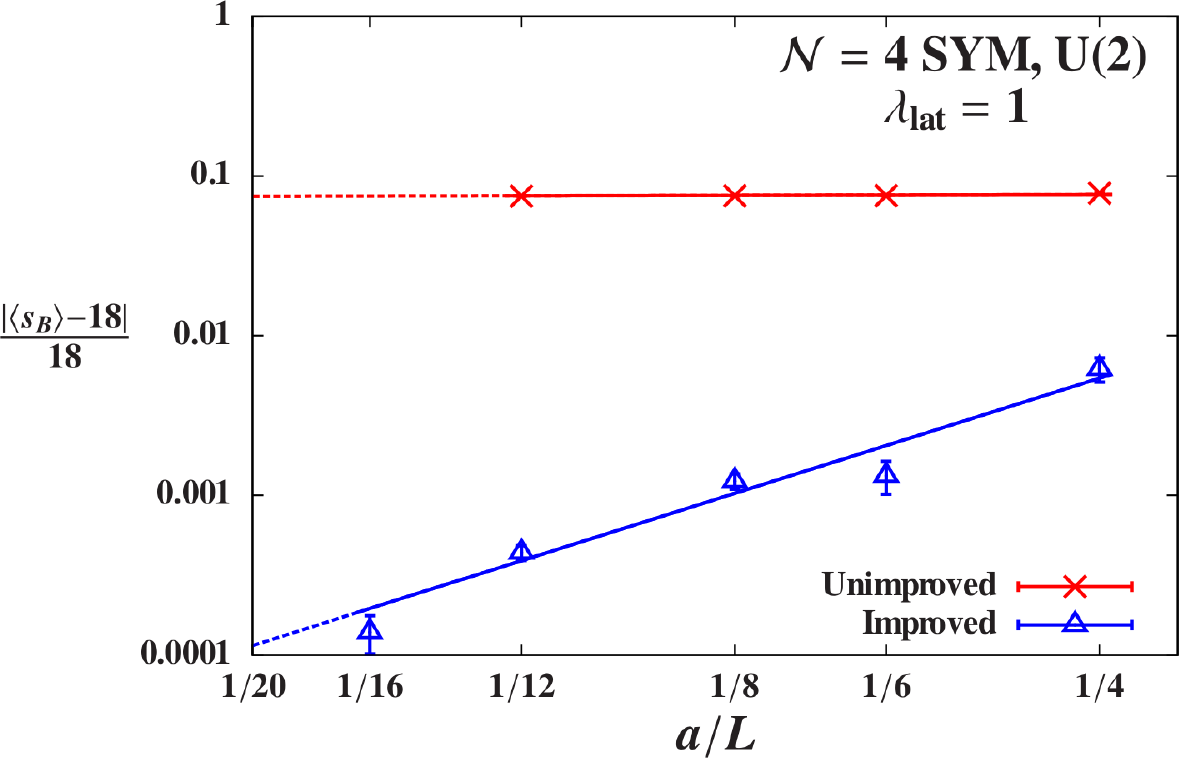}\hfill
  \includegraphics[width=0.475\linewidth]{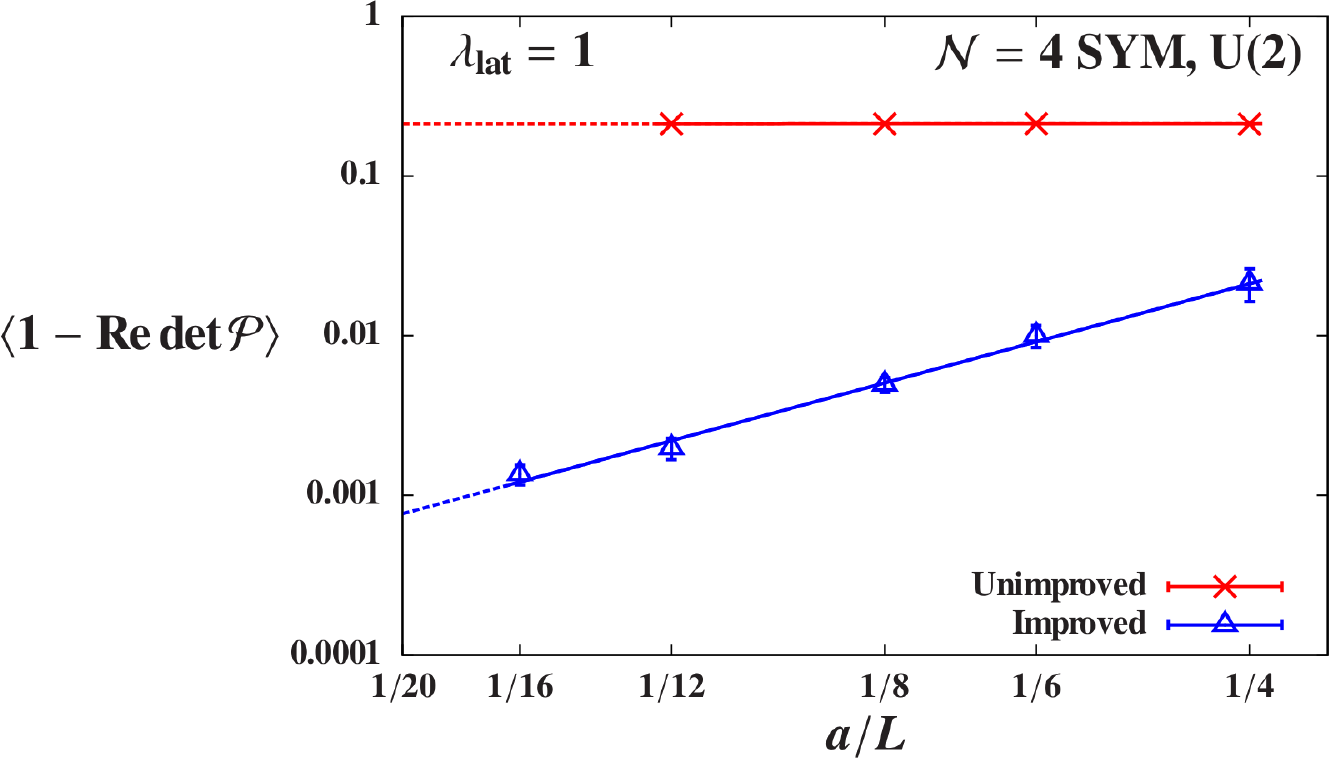}
  \caption{\label{fig:continuum}Continuum extrapolations of \cQ Ward identity violations on log--log axes with power-law fits, for both the improved (blue triangles, Table~\protect\ref{tab:continuum}) and unimproved (red $\times$s) lattice actions with $N = 2$, $\lalat = 1$ and $\mu \sim 1 / L$.  {\bf Left:} Deviations of the bosonic action from its exact supersymmetric value $9N^2 / 2$.  {\bf Right:} Deviations of $\vev{\Re\det\cP}$ from unity.  In both cases the improved action produces much smaller Ward identity violations that vanish $\propto (a / L)^2$ in the continuum limit.}
\end{figure}

Table~\ref{tab:continuum} and \fig{fig:continuum} illustrate how these quantities behave upon approaching the continuum limit with $N = 2$ and $\lalat = 1$.
Violations of each Ward identity appear in the table as deviations from unity, which in all three cases vanish $\propto (a / L)^2$ in the continuum limit, consistent with an $\cO(a)$-improved lattice action.
Indeed, the lattice symmetries (including the \cQ supersymmetry) forbid all dimension-5 operators~\cite{Catterall:2015ira}, so we expect $\cO(a)$ improvement when the soft supersymmetry breaking is sufficiently small.
As mentioned above, the right panel of \fig{fig:continuum} demonstrates that the U(1) sector decouples in the continuum limit ($\Re\det\cP \to 1$) even though $G = 0.05$ is fixed.
In addition, both the Polyakov loop and plaquette are near unity after being normalized by $N$, providing further checks of the individual ensembles.
It is curious that the $\de\cQ\cO$ involving the $\eta \psi_a$ fermion bilinear exhibits significantly larger Ward identity violations than the other two quantities, especially on small lattice volumes.
This pattern persists with either periodic or anti-periodic (thermal) temporal boundary conditions (BCs) for the fermions; all results shown here use thermal BCs.

\begin{table}[htp]
  \caption{\label{tab:4nt4} Some observables from U(2) $4^4$ ensembles with $G = 0.05$ and $(\mu L)^2 / \lalat \approx 2.5$ for $\lalat < 5$.}
  \centering
  \renewcommand\arraystretch{1.2}  
  \begin{tabular}{|ccc|ccccc|}
    \hline
    \lalat  & $\mu$ & $G$   & Plaq.       & $|L|$     & $s_B / 18$    & $\Re\det\cP$  & $1 - \de\cQ\cO$ \\
    \hline
    0.5     & 0.28  & 0.05  & 1.0911(184) & 1.237(43) & 0.99563(99)   & 0.9928(25)    & 0.887(26)       \\
    1.0     & 0.40  & 0.05  & 1.0780(91)  & 1.118(33) & 0.99381(107)  & 0.9788(49)    & 0.885(24)       \\
    2.0     & 0.57  & 0.05  & 1.0638(28)  & 1.066(24) & 0.99060(69)   & 0.9347(35)    & 0.817(11)       \\
    3.0     & 0.69  & 0.05  & 1.0583(35)  & 1.037(23) & 0.98474(58)   & 0.8733(33)    & 0.741(8)        \\
    4.0     & 0.80  & 0.05  & 1.0383(34)  & 0.927(20) & 0.97872(59)   & 0.7975(47)    & 0.668(7)        \\
    \hline
    5.0     & 0.80  & 0.10  & 1.1594(25)  & 1.025(25) & 0.97499(36)   & 0.9418(30)    & 0.819(8)        \\
    6.0     & 0.80  & 0.10  & 1.1974(30)  & 1.156(32) & 0.97301(55)   & 0.9297(32)    & 0.807(8)        \\
    7.0     & 1.00  & 0.15  & 1.2216(23)  & 0.901(36) & 0.95725(62)   & 0.9506(19)    & 0.795(7)        \\
    7.0     & 1.20  & 0.10  & 1.1392(18)  & 0.747(26) & 0.94386(52)   & 0.8416(30)    & 0.590(8)        \\
    8.0     & 1.00  & 0.25  & 1.2805(32)  & 0.897(59) & 0.95134(33)   & 0.9782(17)    & 0.836(10)       \\
    8.0     & 1.20  & 0.15  & 1.2287(18)  & 0.855(24) & 0.93770(42)   & 0.9224(22)    & 0.703(8)        \\
    \hline
  \end{tabular}
\end{table}

\newpage 
Finally in Table~\ref{tab:4nt4} we consider stronger couplings $\lalat \leq 8$ on a fixed $4^4$ lattice volume.
For $\lalat < 5$ we are able to proceed as discussed above, fixing $G = 0.05$ and $(\mu L)^2 / \lalat \approx 2.5$.
However, the corresponding $\lalat = 5$ run with $\mu = 0.89$ is unstable.
In order to reach stronger couplings we either have to increase $G$ or move to larger $L$.
With the same $G = 0.05$ and $(\mu L)^2 / \lalat \approx 2.5$, calculations on $6^4$ lattices encounter no difficulties for $\lalat < 7$.
The bottom half of the table shows some initial $4^4$ explorations of larger $G$, which allow somewhat smaller $\mu$.
Two ensembles with different balances of $(\mu, G)$ are shown for each of $\lalat = 7$ and 8.
Although larger $G$ and smaller $\mu$ lead to better Ward identity violations in both cases, they also produce move severe fluctuations in the pfaffian phase (a worse sign problem)~\cite{Schaich:2015daa}.

\section*{Next steps for lattice $\cN = 4$ SYM} 
Large-scale numerical calculations are now underway using the improved lattice action summarized above, guided by our current understanding of its parameter space.
The primary targets of these studies are the coupling dependence of the Coulomb coefficient $C(\la)$ in the $\cN = 4$ SYM static potential $V(r) = A + C / r$, and the anomalous dimension of the Konishi operator.
In addition we are exploring the possible sign problem of the lattice theory, the restoration of the other supersymmetries $\cQ_a$ and $\cQ_{ab}$ in the continuum limit, and other interesting aspects of the system.
We look forward to reporting initial results from these wide-ranging investigations in the near future.

\vspace{12 pt}
\noindent {\sc Acknowledgments:}~I thank my collaborators on lattice $\cN = 4$ SYM: Simon Catterall, Poul Damgaard, Tom DeGrand and Joel Giedt.
I am also grateful for useful discussions of the lattice parameter space with Anosh Joseph.
Part of this work was performed at the Aspen Center for Physics (U.S.~National Science Foundation [NSF] Grant No.~1066293), the Kavli Institute for Theoretical Physics (NSF Grant No.~PHY11-25915) and Humboldt University (the Emmy Noether Research Group ``Gauge Fields from Strings''), which I thank for their support and hospitality.
Further support came from the U.S.~Department of Energy (DOE), Office of Science, Office of High Energy Physics, under Award Numbers DE-SC0008669 and DE-SC0009998.
Numerical calculations were carried out on the HEP-TH cluster at the University of Colorado and on the DOE-funded USQCD facilities at Fermilab.

\bibliographystyle{utphys}
\bibliography{lattice15}

\providecommand{\href}[2]{#2}\begingroup\raggedright\begin{thebibliography}{10}

\bibitem{Kaplan:2005ta}
D.~B. Kaplan and M.~{\"U}nsal,
  \href{http://dx.doi.org/10.1088/1126-6708/2005/09/042}{{\em JHEP} {\bf 0509}
  (2005) 042} [\href{http://arxiv.org/abs/hep-lat/0503039}{{\tt
  hep-lat/0503039}}].

\bibitem{Unsal:2006qp}
M.~{\"U}nsal, \href{http://dx.doi.org/10.1088/1126-6708/2006/10/089}{{\em
  JHEP} {\bf 0610} (2006) 089}
  [\href{http://arxiv.org/abs/hep-th/0603046}{{\tt hep-th/0603046}}].

\bibitem{Catterall:2007kn}
S.~Catterall,
  \href{http://dx.doi.org/10.1088/1126-6708/2008/01/048}{{\em JHEP} {\bf 0801}
  (2008) 048} [\href{http://arxiv.org/abs/0712.2532}{{\tt arXiv:0712.2532}}].

\bibitem{Damgaard:2008pa}
P.~H. Damgaard and S.~Matsuura,
  \href{http://dx.doi.org/10.1016/j.physletb.2008.01.044}{{\em Phys. Lett.}
  {\bf B661} (2008) 52--56} [\href{http://arxiv.org/abs/0801.2936}{{\tt
  arXiv:0801.2936}}].

\bibitem{Catterall:2009it}
S.~Catterall, D.~B. Kaplan and M.~Unsal,
  \href{http://dx.doi.org/10.1016/j.physrep.2009.09.001}{{\em Phys. Rept.} {\bf
  484} (2009) 71--130} [\href{http://arxiv.org/abs/0903.4881}{{\tt
  arXiv:0903.4881}}].

\bibitem{Ishii:2008ib}
T.~Ishii, G.~Ishiki, S.~Shimasaki and A.~Tsuchiya,
  \href{http://dx.doi.org/10.1103/PhysRevD.78.106001}{{\em Phys. Rev.} {\bf
  D78} (2008) 106001} [\href{http://arxiv.org/abs/0807.2352}{{\tt
  arXiv:0807.2352}}].

\bibitem{Ishiki:2008te}
G.~Ishiki, S.-W. Kim, J.~Nishimura and A.~Tsuchiya,
  \href{http://dx.doi.org/10.1103/PhysRevLett.102.111601}{{\em Phys. Rev.
  Lett.} {\bf 102} (2009) 111601} [\href{http://arxiv.org/abs/0810.2884}{{\tt
  arXiv:0810.2884}}].

\bibitem{Ishiki:2009sg}
G.~Ishiki, S.-W. Kim, J.~Nishimura and A.~Tsuchiya,
  \href{http://dx.doi.org/10.1088/1126-6708/2009/09/029}{{\em JHEP} {\bf 0909}
  (2009) 029} [\href{http://arxiv.org/abs/0907.1488}{{\tt arXiv:0907.1488}}].

\bibitem{Hanada:2010kt}
M.~Hanada, S.~Matsuura and F.~Sugino,
  \href{http://dx.doi.org/10.1143/PTP.126.597}{{\em Prog. Theor. Phys.} {\bf
  126} (2011) 597--611} [\href{http://arxiv.org/abs/1004.5513}{{\tt
  arXiv:1004.5513}}].

\bibitem{Honda:2011qk}
M.~Honda, G.~Ishiki, J.~Nishimura and A.~Tsuchiya,
  \href{http://pos.sissa.it/archive/conferences/139/244/Lattice
  2011_244.pdf}{{\em PoS} {\bf Lattice 2011} (2011) 244}
  [\href{http://arxiv.org/abs/1112.4274}{{\tt arXiv:1112.4274}}].

\bibitem{Honda:2013nfa}
M.~Honda, G.~Ishiki, S.-W. Kim, J.~Nishimura and A.~Tsuchiya,
  \href{http://dx.doi.org/10.1007/JHEP11(2013)200}{{\em JHEP} {\bf 1311} (2013)
  200} [\href{http://arxiv.org/abs/1308.3525}{{\tt arXiv:1308.3525}}].

\bibitem{Hanada:2013rga}
M.~Hanada, Y.~Hyakutake, G.~Ishiki and J.~Nishimura,
  \href{http://dx.doi.org/10.1126/science.1250122}{{\em Science} {\bf 344}
  (2014) 882--885} [\href{http://arxiv.org/abs/1311.5607}{{\tt
  arXiv:1311.5607}}].

\bibitem{Catterall:2014vka}
S.~Catterall, D.~Schaich, P.~H. Damgaard, T.~DeGrand and J.~Giedt,
  ``{$\mathcal N = 4$ supersymmetry on a space-time lattice}'',
  \href{http://dx.doi.org/10.1103/PhysRevD.90.065013}{{\em Phys. Rev.} {\bf
  D90} (2014) 065013} [\href{http://arxiv.org/abs/1405.0644}{{\tt
  arXiv:1405.0644}}].

\bibitem{Catterall:2014vga}
S.~Catterall, J.~Giedt, D.~Schaich, P.~H. Damgaard and T.~DeGrand, ``{Results
  from lattice simulations of $\mathcal N = 4$ supersymmetric Yang--Mills}'',
  \href{http://pos.sissa.it/archive/conferences/214/267/LATTICE2014_267.pdf}{{\em PoS}
  {\bf LATTICE2014} (2014) 267}
  [\href{http://arxiv.org/abs/1411.0166}{{\tt arXiv:1411.0166}}].

\bibitem{Catterall:2015ira}
S.~Catterall and D.~Schaich, ``{Lifting flat directions in lattice
  supersymmetry}'', \href{http://dx.doi.org/10.1007/JHEP07(2015)057}{{\em JHEP}
  {\bf 07} (2015) 057}, [\href{http://arxiv.org/abs/1505.03135}{{\tt
  arXiv:1505.03135}}].

\bibitem{Schaich:2015daa}
D.~Schaich and S.~Catterall, ``{Maximally supersymmetric Yang-Mills on the
  lattice}'', in {\em {Sakata Memorial KMI Workshop on Origin of Mass and
  Strong Coupling Gauge Theories (SCGT15), Nagoya, 3--6 March}} (2015)
  [\href{http://arxiv.org/abs/1508.00884}{{\tt arXiv:1508.00884}}].

\bibitem{Schaich:2014pda}
D.~Schaich and T.~DeGrand, ``{Parallel software for lattice $\mathcal N = 4$
  supersymmetric Yang--Mills theory}'',
  \href{http://dx.doi.org/10.1016/j.cpc.2014.12.025}{{\em Comput. Phys.
  Commun.} {\bf 190} (2015) 200--212}
  [\href{http://arxiv.org/abs/1410.6971}{{\tt arXiv:1410.6971}}].

\bibitem{Catterall:2014mha}
S.~Catterall and J.~Giedt,
  \href{http://dx.doi.org/10.1007/JHEP11(2014)050}{{\em JHEP} {\bf 1411} (2014)
  050} [\href{http://arxiv.org/abs/1408.7067}{{\tt arXiv:1408.7067}}].

\bibitem{Marcus:1995mq}
N.~Marcus, \href{http://dx.doi.org/10.1016/0550-3213(95)00389-A}{{\em
  Nucl. Phys.} {\bf B452} (1995) 331--345}
  [\href{http://arxiv.org/abs/hep-th/9506002}{{\tt hep-th/9506002}}].

\bibitem{Kapustin:2006pk}
A.~Kapustin and E.~Witten,
  \href{http://dx.doi.org/10.4310/CNTP.2007.v1.n1.a1}{{\em Commun. Num. Theor.
  Phys.} {\bf 1} (2007) 1--236}
  [\href{http://arxiv.org/abs/hep-th/0604151}{{\tt hep-th/0604151}}].

\bibitem{Catterall:2011pd}
S.~Catterall, E.~Dzienkowski, J.~Giedt, A.~Joseph and R.~Wells,
  \href{http://dx.doi.org/10.1007/JHEP04(2011)074}{{\em JHEP} {\bf
  1104} (2011) 074} [\href{http://arxiv.org/abs/1102.1725}{{\tt
  arXiv:1102.1725}}].

\end{thebibliography}\endgroup
\end{document}